\documentclass[finalprint, 5p]{elsarticle}
\usepackage{graphicx}
\usepackage{amsmath}
\usepackage{makeidx}
\usepackage{amssymb}
\usepackage{float}
 
\begin{document}

\begin{frontmatter}

\title{Thomas Fermi approximation and large-$N$ quantum mechanics}
\author[label1]{Sukla Pal\corref{cor1}}
\ead{sukla@bose.res.in}
\author[label2]{Jayanta K. Bhattacharjee}
\address[label1]{Department of Theoretical Physics, S.N.Bose National Centre For Basic Sciences, JD-Block, Sector-III, Salt Lake City, Kolkata-700098, India}
\address[label2]{Harish-Chandra Research Institute, Chhatnag road, Jhunsi, Allahabad-211019, India}
\cortext[cor1]{Corresponding author}
\begin{abstract}
We note that the Thomas Fermi limit of Gross Pitaevskii equation and $N>>1$ limit of quantum mechanics, where $N$ is the dimensionality of space, are based on the same point of view. We combine these two to produce a modified Thomas Fermi approximation which gives a very good account of the energy of the condensate in harmonic trap.
\end{abstract}
\end{frontmatter}
\section{Introduction}Bose Einstein condensation has been experimentally achieved with particles in a trap-generally a simple harmonic oscillator potential. The condensate is well described in terms of the Gross Pitaevskii equation (GPE)\cite{1,2,3,4} both for the energy (chemical potential) of the stationary state and as well as dynamics. To find the lowest energy of the system, one can use a variational estimate with a Gaussian trial function for a Thomas Fermi approximation which ignores the kinetic energy of the particles. The Thomas Fermi approximation works well for a large number of particles but generally is an underestimate as compared to the variational calculation. Hence finding the correction to the Thomas Fermi limit is of interest. The first attempt at doing this is the work of Schuch and Vin \cite{5} who combined a WKB type of analysis with the Thomas Fermi idea to obtain a very effective modified Thomas Fermi energy. It has not been noticed yet that analogous to the Thomas Fermi limit is the infinite dimension (large $N$) limit for finding the ground state energy of a quantum mechanical problem \cite{6,7,8,9}. As $N\rightarrow\infty$, the kinetic energy becomes negligible and one finds the ground state energy from a minimization of an effective potential. Corrections are obtained in a systematic fashion by first considering simple harmonic motion about the minimum and then including anharmonicity. Our point in this article is that the first correction to the Thomas Fermi energy can be found in a manner analogous to the correction to the $N\rightarrow\infty$ limit and also the large $N$ approach can be combined with the Thomas Fermi limit to yield accurate answers more easily. 

In sec II, we will recall the large $N$ quantum mechanical problem. We will first consider the harmonic oscillator potential where the exact answer is obtained at the leading order itself and all subsequent corrections are zero. More pertinent is the an-harmonic oscillator which we treat next and show that a two term answer is a significant improvement on the leading terms. In section III, we use the philosophy of approach to find the correction to the Thomas Fermi limit in a one dimensional problem. In sec IV, we combine the large $N$ approach with the Thomas Fermi limit to obtain the large $N$ approach with the Thomas Fermi limit to obtain a very reasonable estimate of the stationary state energy.

\section{1/N expansion}  
In this section, we recall the $1/N$ expansion for the harmonic oscillator and the an-harmonic oscillator. We will get the ground state energy to two term accurately in each case and compare with the known exact (analytic or numerical) answer at $N=1$, which is the worst situation for the technique.Even in this very unfavorable situation, the comparison between approximation and reality is good, giving us confidence in this procedure.

\subsection{Harmonic oscillator potential}
The laplacian operator for the zero angular momentum state in a N-dimensional space has the form 
\begin{equation}\label{eq1}
\nabla^2=\frac{1}{r^{N-1}}\frac{d}{dr}(r^{N-1}\frac{d}{dr})
\end{equation}
Consequently, the Schr$\ddot{o}$dinger equation for the ground state of the simple harmonic oscillator of frequency $\omega$ reads
\begin{equation}\label{eq2}
-\frac{\hbar^2}{2m}\frac{1}{r^{N-1}}\frac{d}{dr}(r^{N-1}\frac{d\phi}{dr})+\frac{1}{2}m\omega^2r^2\phi=E\phi
\end{equation}
where, $m$ is the mass of the particle and $E$ is the ground state energy. If we carry out the transformation $\phi=\frac{u}{r^{\frac{N-1}{2}}}$, then
\begin{eqnarray}\label{eq3}
\begin{aligned}
Hu&=-\frac{\hbar^2}{2m}[\frac{d^2u}{dr^2}-\frac{(N-1)(N-3)}{4r^2}u]+\frac{1}{2}m\omega^2r^2u\\&=Eu
\end{aligned}
\end{eqnarray}
where, $`H'$ is the Hamiltonian operator. We define a dimensionless spatial variable $\xi$ as 
\begin{equation}\label{eq4}
\xi^2=r^2/\frac{\hbar}{m\omega}N
\end{equation}
so that the Hamiltonian now turns out to be 
\begin{equation}\label{eq5}
\frac{H}{\hbar\omega}=N[-\frac{1}{2N^2}\frac{d^2}{d\xi^2}+\frac{(1-\frac{1}{N})(1-\frac{3}{N})}{8\xi^2}+\frac{\hbar\omega}{2}\xi^2]
\end{equation}
Rescalling $E=N\hbar\omega\epsilon$, we can rewrite Eq. (\ref{eq3})
\begin{eqnarray}\label{eq6}
\begin{aligned}
-\frac{1}{2N^2}\frac{d^2}{d\xi^2}+\frac{(1-\frac{1}{N})(1-\frac{3}{N})}{8\xi^2}+\frac{\hbar\omega}{2}\xi^2&=\frac{E}{N\hbar\omega}u\\&=\epsilon u
\end{aligned}
\end{eqnarray}
As $N\rightarrow\infty$, only the $O(1)$ terms on the right hand side of Eq. (\ref{eq6}) survive and in the ground state the particle sets at the minimum of the effective potential $V=\frac{1}{8\xi^2}+\frac{\xi^2}{2}$. The minimum is at $\xi=\xi_0$ with $\xi_0^2=1/2$ and in this approximation the ground state energy is $\epsilon_0=1/2$ and given by 
\begin{equation}\label{eq7}
E_0=N\hbar\omega\epsilon_0=\frac{N\hbar\omega}{2}
\end{equation}
This is what in critical phenomena would be called the spherical limit \cite{10,11}. In this context it is the leading term for $N>>1$. The first correction to the leading order is obtained by expanding $V$ about $\xi_0$ and writing $H=\frac{1}{2}+H'$, where $\frac{1}{2}$ is the $N\rightarrow\infty$ form and for finite $N$, $H'$ is the variation around the infinite $N$ limit. To the leading order, $H'$ can be expressed by expanding around $\xi_0=1/2$
\begin{equation}\label{eq8}
H'=-\frac{1}{2N^2}\frac{d^2}{d\xi^2}+2(\xi-\xi_0)^2-\frac{1}{N}
\end{equation}
which is the Hamiltonian of simple harmonic oscillator of frequency $\omega'$ with $\omega'^2=4/d^2$ and an additional part. The ground state energy is clearly $\frac{1}{2}\frac{2}{N}-\frac{1}{N}=0$ and hence there is no correction at this order. The ground state energy thus remains $E_0=\frac{N\hbar\omega}{2}+O(\frac{1}{N})$ which is comparable to the exact answer of $\frac{N}{2}\hbar\omega$. Higher order corrections will remain zero because in this case $\frac{N\hbar\omega}{2}$ is the exact answer.

\subsection{Anharmonic oscillator ($\lambda x^4$ potential)}
Had the potential been $\lambda x^4$ in place of $\frac{1}{2}m\omega^2x^2$, we would have been able to formulate an identical scheme of expansion and corrections could no longer be zero. In this case, Eq. (\ref{eq3}) would take the form
\begin{eqnarray}\label{eq9}
\begin{aligned}
Hu&=-\frac{\hbar^2}{2m}[\frac{d^2u}{dr^2}-\frac{N^2(1-\frac{1}{N})(1-\frac{3}{N})}{4r^2}u]+\frac{\lambda}{4}r^4u\\
&=Eu
\end{aligned}
\end{eqnarray}
Considering the scaling as
\begin{eqnarray}
r&=&(\frac{\hbar^2}{m\lambda})^{1/6}N^{1/3}\xi\label{eq10}\\
E&=&N^{4/3}(\frac{\hbar^2}{m})^{2/3}\lambda^{1/3}\epsilon\label{eq11}
\end{eqnarray}
Eq. (\ref{eq9}) reduces to
\begin{eqnarray}\label{eq12}
Hu=-\frac{1}{2N^2}[\frac{d^2u}{d\xi^2}-\frac{(1-\frac{1}{N})(1-\frac{3}{N})}{8\xi^2}u]+\frac{\xi^4}{4}u=\epsilon u
\end{eqnarray}
In the limit of $N\rightarrow\infty$, there is only the effective potential $V=\frac{1}{8\xi^2}+\frac{\xi^4}{4}$ and for the ground state, the particle sits in the minimum of the potential which occurs at $\xi=\xi_0$, where $\xi^6=4^{-1}$ and $V_{min}=\frac{3}{16}\times4^{1/3}$ giving $\epsilon_0=\frac{3}{16}\times4^{1/3}$.

To obtain the first correction $H'$, we need to keep the derivative term in the Hamiltonian of Eq. (\ref{eq12}). Keeping the term $-\frac{1}{2N\xi_0^2}$ in $H_0$ and expanding the effective potential $V$ about $\xi_0$ to quadratic order we get
\begin{equation*}
V\simeq\frac{3}{16}\times4^{1/3}+2.45(\Delta\xi)^2
\end{equation*}
In this approximation
\begin{equation*}
H'=-\frac{1}{2N^2}\frac{d^2u}{d(\Delta\xi)^2}+2.45(\Delta\xi)^2-\frac{4^{1/3}}{2N}
\end{equation*}
The ground state energy $\epsilon_1$ of $H'$ is 
\begin{equation*}
\epsilon_1=\frac{1}{2}\sqrt{\frac{4.9}{N^2}}-\frac{4^{1/3}}{2N}\simeq\frac{0.31}{N}
\end{equation*}
To this order the energy $E$ is given by 
\begin{equation}\label{eq13}
E=(\frac{\hbar^2}{m})^{2/3}\lambda^{1/3}N^{4/3}[\frac{3}{16}\times4^{1/3}+\frac{0.31}{N}+O(\frac{1}{N^2})]
\end{equation}
For $N=1$, at this order we get $E\simeq0.61(\frac{\hbar^2}{m})^{2/3}\lambda^{1/3}$, which is to be compared with the numerically exact value of  $E\simeq0.668(\frac{\hbar^2}{m})^{2/3}\lambda^{1/3}$. Considering the fact that $N=1$ as far remote form of $N\rightarrow\infty$ as possible and we have calculated to a reasonably trivial order, this agreement is quite impressive. The agreement of the two term answer above is far more impressive for $N=3$. In the next section, we consider the Gross Pitaevskii model
\section{Gross Pitaevskii Model with a harmonic trap in 1N}
For the N particle Bose Einstein condensate in a harmonic trap in 1N, the wave function `$u$' satisfies the Gross-Pitaevskii equation (GPE) 
\begin{eqnarray}\label{eq14}
-\frac{\hbar^2}{2m}\frac{d^2u}{dx^2}+\frac{1}{2}m\omega^2x^2u+g|u|^2u=Eu
\end{eqnarray}
where, the coupling constant $`g'$ corresponds to the two body interaction between the particles that constitutes the condensate. For $g>0$, it signifies repulsive interaction while for $g<0$, attractive interaction is indicated. In this work we will be concerned with $g>0$. We carry out the following rescalings,
\begin{eqnarray}\label{eq15}
\begin{aligned}
x&=\sqrt{\frac{\hbar}{m\omega}}\xi\\
u&=(\frac{m\omega}{\hbar})^{1/4}\phi\\
g&=\lambda\hbar\omega\sqrt{\frac{\hbar}{m\omega}}
\end{aligned}  
\end{eqnarray}
to arrive at the following
\begin{equation}\label{eq16}
-\frac{1}{2}\frac{d^2\phi}{d\xi^2}+\frac{\xi^2}{2}\phi+\lambda|\phi|^2\phi=\frac{E}{\hbar\omega}\phi
\end{equation}
The scaling given in Eq. (\ref{eq15}), ensures that $\int_{\infty}^{\infty}|\phi|^2d\xi=1$. 
\subsection{Energy obtained from variational calculation:}
Considering the trial wave function of the form, $\phi(x)=\frac{1}{\Delta^{1/2}\pi^{1/4}}e^{-\xi^2/2\Delta^2}$ the energy of the 1N condensate is expressed as
\begin{equation}\label{eq17}
\frac{E}{\hbar\omega}=\frac{1}{4}(\frac{1}{\Delta^2}+\Delta^2)+\frac{\lambda}{\sqrt{2\pi}}\frac{1}{\Delta}
\end{equation}
Minimizing with respect to $\Delta$, we get the value $\Delta_0$ at the minimum to satisfy
\begin{equation}\label{eq18}
\frac{1}{2\Delta_0^3}+\frac{\lambda}{\sqrt{2\pi}\Delta_0^2}=\frac{\Delta_0}{2}
\end{equation}
Obtaining $\Delta_0$ for different values of $\lambda$, we can find the ground state energy from Eq. (\ref{eq17}) as a function of $\lambda$ and this is plotted as the solid curve in Fig. \ref{fig2}.
\subsection{Energy obtained from large $\lambda$ expansion}
To explore the large $\lambda$ behavior we consider the scale transformation $\xi=\lambda^{1/2}y$
\begin{equation}\label{eq19}
-\frac{1}{2\lambda^2}\frac{d^2\phi}{dy^2}+\frac{y^2}{2}\phi+|\phi|^2\phi=\epsilon\phi
\end{equation}
where $E=\epsilon\lambda\hbar\omega$. We would like to compare the above with Eq. (\ref{eq6}). The present situation is more complicated since instead of a term proportional to $\xi^{-2}$ we now have $|\phi|^2$ which can be known only after the problem is solved. However we note that $|\phi|^2\rightarrow 0$ for $\xi\gg1$ and hence the two terms are of the same character. We try to exploit this analogy. 

We take the limit of $\lambda\rightarrow\infty$ and drop the first term in Eq. (\ref{eq16}) to write the solution $\phi=\phi_0(y)$ at this order as 

\begin{eqnarray}\label{eq20}
\begin{aligned}
|\phi_0|^2 &=\epsilon_0-\frac{y^2}{2}~~~~~~\text{for $y^2\le y_m^2=2\epsilon_0$}\\
           &=0 ~~~~~~~~~~~~~~~~~~~{\text{otherwise}}
\end{aligned}
\end{eqnarray}
This is the Thomas Fermi approximation and the value of $\epsilon_0$ is fixed by the normalization condition $\frac{1}{\lambda^{1/2}}=\int|\phi|_0^2dy=\frac{2}{3}(2\epsilon_0)^{3/2}$ which leads to 
\begin{equation}\label{eq21}
\epsilon_0=\frac{1}{2}(\frac{3}{2\lambda^{1/2}})^{2/3}
\end{equation}

\begin{figure}[H]
\includegraphics[angle=0,scale=0.8]{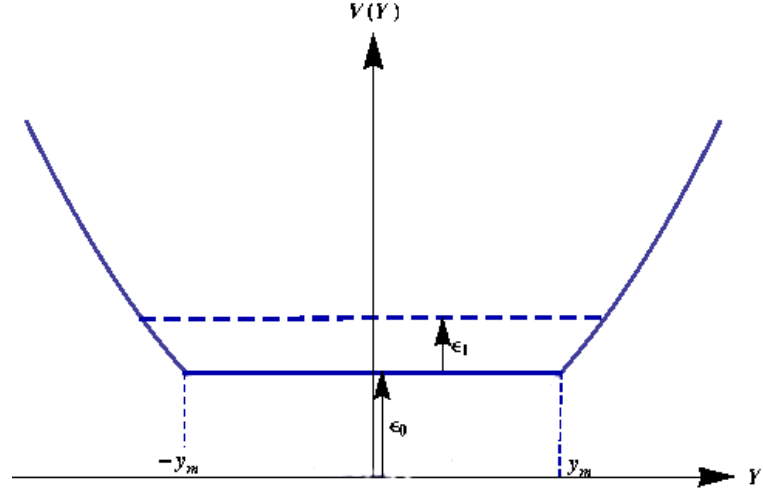}
\caption{Above figure describes the schematic diagram for the potential given in Eq. (\ref{eq23}). All other parameters that we have considered in the further analysis of WKB have also been shown in this figure.}
\label{fig1}
\end{figure}

\begin{figure}[H]
\includegraphics[angle=0,scale=0.7]{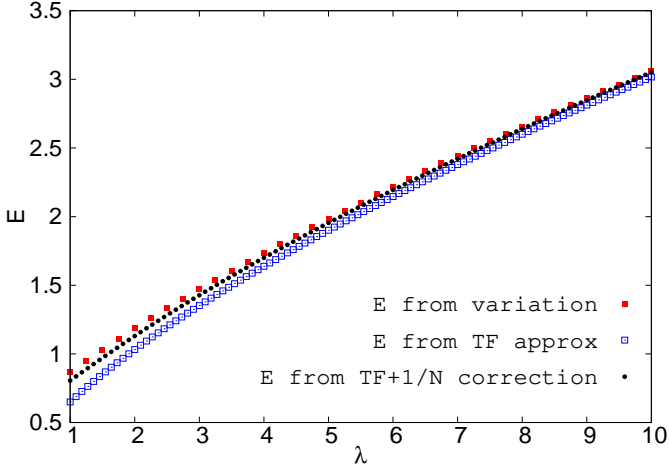}
\caption{Black curve shows the ground state energy of IN GPE, obtained after including the first order correction over the usual Thomas Fermi approximation. Variational result is indicated by the red curve and the blue curve shows the energy values obtained from TF approximation.}
\label{fig2}
\end{figure}

If we view the Hamiltonian as $H=-\frac{1}{2\lambda^2}\frac{d^2}{dy^2}+\frac{y^2}{2}+|\phi|^2$, then we have taken into account the part of $H$ which is $\frac{y^2}{2}+|\phi|^2$ and this extends over the region $|y|\le\sqrt{2\epsilon_0}$. The remainder of the Hamiltonian is $-\frac{1}{2\lambda^2}\frac{d^2}{dy^2}$ over all space and a potential which is $\frac{y^2}{2}$ for $|y|\ge\sqrt{2\epsilon_0}$ and $\epsilon_0$ for $|y|\le\sqrt{2\epsilon_0}$. The lowest energy eigen value of the potential is $\epsilon_1$ relative to a base value of $\epsilon_0$. and the corresponding V(y) is shown in Fig. \ref{fig1}. Hence we need to solve 
\begin{equation}\label{eq22} 
-\frac{1}{2\lambda^2}\frac{d^2\phi}{dy^2}+V(y)\phi=(\epsilon_0+\epsilon_1)\phi
\end{equation}
With the analytic form of $V(y)$, as
\begin{eqnarray}\label{eq23}
\begin{aligned}
V(y)&=\frac{y^2}{2}~~~\text{for $|y|\ge\sqrt{2\epsilon_0}$}\\
           &=\epsilon_0~~~~\text{for $|y|\le\sqrt{2\epsilon_0}$}
\end{aligned}
\end{eqnarray}
It will be very cumbersome to find the exact answer (which can be done by matching trigonometric functions and Weber functions at $y=\pm\sqrt{2\epsilon_0}$) and so we resort to a WKB procedure which for the lowest value of $\epsilon_1$ yields
\begin{eqnarray}\label{eq24}
\begin{aligned}
2\lambda&\big[\int_0^{y_m}\sqrt{2(\epsilon_1+\epsilon_0-V(y))}dy+\\&\int_{y_m}^{y_0}\sqrt{2(\epsilon_1+\epsilon_0-\frac{y^2}{2})}dy\big]=\frac{\pi}{2}
\end{aligned}
\end{eqnarray}
Where $y_0$ is to be obtained from $y_0^2=2(\epsilon_0+\epsilon_1)$. The first integral on the left hand side of Eq. \ref{eq24} is easily seen to be $2\sqrt{\epsilon_0\epsilon_1}$, while the second integral is evaluated to the leading order in $\sqrt{\frac{\epsilon_1}{\epsilon_0}}$. We have
\begin{eqnarray}\label{eq25}
\int_0^{y_0}\sqrt{2(\epsilon_1+\epsilon_0)-y^2}dy=(\epsilon_1+\epsilon_0)[\theta_0-\frac{\sqrt{\epsilon_1\epsilon_0}}{\epsilon_0+\epsilon_1}]
\end{eqnarray}
where, $Cos^2\theta_0=\frac{\epsilon_0}{\epsilon_0+\epsilon_1}$. Extracting $\theta_0$ to O($\sqrt{\frac{\epsilon_1}{\epsilon_0}}$), we get
\begin{eqnarray}
\sqrt{\frac{\epsilon_1}{\epsilon_0}}&=&\frac{\pi}{4}(\sqrt{2}-1)\frac{1}{\lambda\epsilon_0}\nonumber\\
&=&\frac{\pi}{2}(\sqrt{2}-1)(\frac{2}{3})^{2/3}\frac{1}{\lambda^{2/3}}\label{eq26}
\end{eqnarray}
We have the two term Thomas Fermi energy consequently given by 
\begin{eqnarray}
E=\frac{\hbar\omega}{2}(\frac{3}{2})^{2/3}\lambda^{2/3}[1+\frac{\pi^2}{4}(\sqrt{2}-1)^{2}(\frac{2}{3})^{4/3}\frac{1}{\lambda^{4/3}}\nonumber\\+\cdots]\label{eq27}
\end{eqnarray}
\section{Combining large N and Thomas Fermi approximation}
In this section, we seek an improvement on the Thomas Fermi technique by starting with the GPE in N dimension and doing simultaneously a large N and large coupling constant approximation. To this end we begin with
\begin{equation}\label{eq28}
-\frac{\hbar^2}{2m}\nabla^2\psi+\frac{1}{2}m\omega^2r^2\psi+g|\psi|^2\psi=E\psi
\end{equation}
as the stationary state GPE and use the transformation $\psi=\frac{u}{r^{(N-1)/2}}$ to write for $N>>1$ as
\begin{equation}\label{eq29}
-\frac{\hbar^2}{2m}[\frac{d^2}{dr^2}-\frac{N^2}{4r^2}]u+\frac{1}{2}m\omega^2r^2u+g\frac{|u|^2u}{r^{N-1}}=Eu
\end{equation}
with $r$ scaled by $\sqrt{\frac{\hbar}{m\omega}}$ (the oscillator length) and we get
\begin{equation}\label{eq30}
-\frac{1}{2}\frac{d^2}{d\xi^2}u+\frac{N}{8\xi^2}u+\frac{\xi^2}{2}u+(\frac{m\omega}{\hbar})^{\frac{N-1}{2}}\frac{g}{\hbar\omega}\frac{|u|^2u}{\xi^{N-1}}=\frac{E}{\hbar\omega}u
\end{equation}
with the normalization condition $\int|u|^2d\xi=\frac{\Gamma(N/2)}{2\pi^{N/2}}$. We now make a further rescalling by $N$ as done in sec II and with $\xi=N^{1/2}R$ we reach into the following
\begin{equation}\label{31}
\begin{aligned}
-\frac{1}{2N^2}\frac{d^2u}{dR^2}+\frac{u}{8R^2}+\frac{R^2}{2}u+&(\frac{m\omega}{\hbar})^{\frac{N-1}{2}}\frac{g}{\hbar\omega N^{\frac{N+1}{2}}}\\&\times\frac{|u|^2}{R^{N-1}}u=\frac{E}{\hbar\omega N}u
\end{aligned}
\end{equation}
In the large $g$, large $N$ limit keeping the ratio $\frac{g}{N^{(N+1)/2}}$ finite, the term O$(\frac{1}{N^2})$ can be dropped and we have the modified Thomas Fermi form for the wave function given by 
\begin{equation}\label{eq32}
(\frac{m\omega}{\hbar})^{\frac{N-1}{2}}\frac{g}{\hbar\omega N^{\frac{N+1}{2}}}\frac{|u|^2}{N^{\frac{N+1}{2}}}=(\epsilon-\frac{R^2}{2}-\frac{1}{8R^2})R^{N-1}
\end{equation}
where $\epsilon=\frac{E}{\hbar\omega N}$ and the normalization now becomes 
\begin{equation}\label{eq33}
\frac{2\pi^{N/2}}{\Gamma(N/2)}\sqrt{\frac{\hbar}{m\omega}}\sqrt{N}\int_{R_1}^{R_2}|u|^2dR=1
\end{equation}
where $R_1$ and $R_2$ are the limits between which $u$ is non zero i.e., $R_1$ and $R_2$ correspond to the zeroes of the right hand side of Eq. (\ref{eq29}) with
\begin{equation}\label{eq34}
R_{2,1}^2=\epsilon\pm\sqrt{\epsilon^2-\frac{1}{4}}
\end{equation}
The normalization condition given in Eq. (\ref{eq33}) becomes with the wave function of Eq. (\ref{eq32})
\begin{equation}\label{eq35}
\begin{aligned}
&\frac{2\pi^{N/2}}{\Gamma(N/2)}(\frac{\hbar}{m\omega})^{N/2}\frac{N^{\frac{N}{2}+1}}{g}\Big[\frac{\epsilon}{N}(R_2^N-R_1^N)-\\&\frac{(R_2^{N+2}-R_1^{N+2})}{2(N+2)}-\frac{(R_2^{N-2}-R_1^{N-2})}{8(N-2)}\Big]=1
\end{aligned}
\end{equation}
Along with the expression for $R_1$ and $R_2$ from Eq. (\ref{eq34}), we have the large N modified Thomas Fermi approximation for $E$ in Eq. (\ref{eq35}). If we now set N=3, after few simple algebra we reach into
\begin{equation}\label{eq36}
\begin{aligned}
\sqrt{2(\frac{E}{\hbar\omega})-3}\Big[\frac{4}{15}(\frac{E}{\hbar\omega})^2+&(\frac{E}{\hbar\omega})\frac{1}{5}-\frac{9}{10}\Big]\\&=\frac{g}{4\pi\hbar\omega}(\frac{m\omega}{\hbar})^{3/2}
\end{aligned}
\end{equation}
If $a$ is the scattering length and $n$ be the total number of particles then $\frac{g}{4\pi\hbar\omega}=\frac{\hbar}{m\omega}an$ and writing $\sqrt{\frac{\hbar}{m\omega}}=a_{osc}$, we can rewrite Eq. (\ref{eq36}) as (with $\bar{E}=\frac{E}{\hbar\omega}$)
\begin{equation}\label{eq37}
\frac{4}{15}(2\bar{E}-3)^{1/2}[\bar{E}^2+\frac{3}{4}\bar{E}-\frac{27}{8}]=\frac{na}{a_{osc}}
\end{equation}
We have found $\bar{E}$ coming from this equation for $Cs^{137}$ atoms for which $a=3\times10^{-7}$ agrees reasonably well to the value of the ground state energy obtained from standard quantum mechanics calculation and  the value at TF Fermi limit. We have shown in Table 1 the change of ground state energy ($\frac{E}{\hbar\omega}$) with the total number of atoms ($n$), $1/N$ correction method having higher accuracy for higher values of $n$.
 
\begin{figure}[H]
\includegraphics[angle=0,scale=0.35]{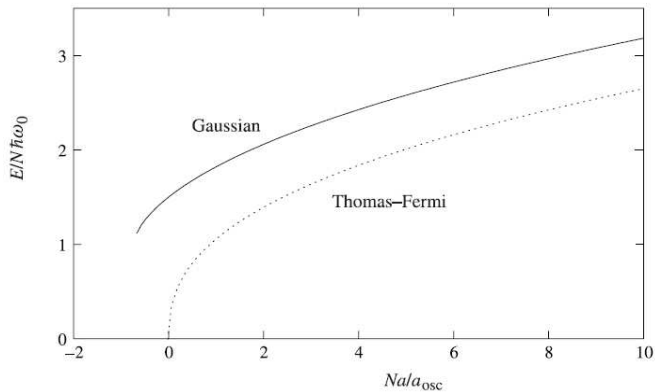}
\caption{This figure has been taken from the book `Bose-Einstein Condensation in Dilute Gases' by C. J. Pethick and H. Smith. (Cambridge university press) chap 6 page no. 155. The figure shows variational estimate of the energy per particle for an isotropic harmonic trap as a function of the dimensionless parameter $Na/a_{osc}$, where '$a$' is scattering length, $a_{osc}$ is oscillator length and for this figure only $N$ indicates the total number of particle i.e., $N=n$. The dotted line is the result obtained from the Thomas Fermi approximation.}
\label{fig3}
\end{figure}
In general the Thomas Fermi answer for the energy of the condensate falls significantly below the variational calculation (generally close to the real answer) for small values of $n$ because of the neglect of kinetic energy term. This is seen from Fig. \ref{fig3} taken from the work by Pethick and Smith \cite{3}.
\begin{figure}[H]
\includegraphics[angle=0,scale=0.7]{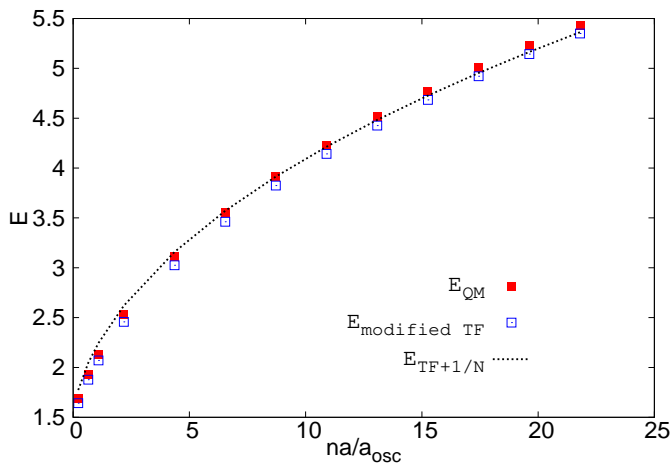}
\caption{Above figure is plotted taking the values (2nd, 3rd and 4th column) of Table 1 with $\frac{na}{a_{osc}}$. This figure clearly signifies the strength of large-N quantum mechanics applied in higher dimension also.}
\label{fig4}
\end{figure}
\begin{table}[H]
 \centering 
\begin{tabular}{c  c  c  c} 
\hline\hline 
$n$ & $E_{TF+1/N}$ & $E_{QM}$ & $E_{TF}$(Modified TF of \cite{5})  \\
\hline 
200 & 1.774 & 1.688 & 1.642\\
600 & 2.046 & 1.927 & 1.877\\
1000 & 2.245 & 2.134 & 2.071\\
2000 & 2.62 & 2.535 & 2.457\\
4000 & 3.159 & 3.112 & 3.025\\
6000 & 3.571 & 3.550 & 3.461\\
8000 & 3.914 & 3.914 & 3.825\\
10000 & 4.214 & 4.231 & 4.142\\ 
12000 & 4.483 & 4.513 & 4.426\\
14000 & 4.727 & 4.770 & 4.684\\
16000 & 4.954 & 5.007 & 4.921\\ 
18000 & 5.165 & 5.228 & 5.143 \\
20000 & 5.363 & 5.435 & 5.350\\
\hline 
\end{tabular}
\caption{Ground state energy is calculated through $1/N$ expansion method in the second column.The frequency of the harmonic oscillator is $\omega=20\pi s^{-1}$ and mass of Cs is taken 133 amu. The energy values given in 3rd (from general quantum mechanical calculation) and 4 th column are taken from Table II of the work by P. SCHUCK and X. VINAS \cite{5} to compare the accuracy of our results. With increase of $n$, $1/N$ method gives quite impressive result than TF limit as expected.}
\end{table}
\section{Conclusion}
In conclusion, we have proposed the method of large-N quantum mechanics and have applied this method to various 1D systems (harmonic oscillator, an-harmonic oscillator, GP model for dilute BEC at $T=0$ both in lower and higher dimension) and have derived the significant correction in the leading order. For all these cases, the corrections have shown sufficient improvement over the base values. In case of GP model, the energy corrections obtained by this method have shown quite a remarkable improvement over the usual Thomas Fermi approximation. Comparison with the numerical values obtained for the ground state energy as given in column 2 of Table 1 justifies the potential of this method.
\section*{Acknowledgments}
One of the authors, Sukla Pal would like to thank S. N. Bose National Centre for Basic Sciences for the financial support during the work. Sukla Pal acknowledges Harish-Chandra Research Institute for hospitality and support during visit.


\begin{thebibliography}{99}
\bibitem{1}F. Dalvano, S. Giorgini, Lev. P. Pitaevskii and S. Stringari, Rev. Mod. Phys. 71 (1999) 463
\bibitem{2}F. Dalfovo, L. Pitaevskii and S. Stringari, Phys. Rev. A,  54 (2000) 4213
\bibitem{3}C. J. Pethick and H. Smith, Bose Einstein condensation in dilute gases, Cambridge University Press, (2002) 154-161
\bibitem{4}S. K. Adhikari and P Muruganandam, J. Phys. B: At. Mol. Opt. Phys. 35 (2002) 2831-43
\bibitem{5}P. Schuck and X. Vin, Phys. Rev. A, 61 (2000) 043603
\bibitem{6}E. Witten, Nuc Phys B, 185 (1981) 513
\bibitem{7}F. Cooper and B. Freedman, Ann. Phys (NY), 146 (1983) 262
\bibitem{8}L. Modinow and N. Papanicolaon, Phys. Rev. A, 25 (1982) 1305
\bibitem{9}T. D. Imbo and U. Sukhatme, Phys. Rev. Lett, 54 (1985) 218
\bibitem{10}S. K. Ma, Phys. Rev. A, 7 (1973) 2172
\bibitem{11}M. Moohe and J. Zinnjistin, Phys. Reports 385 (2003) 69-228
\end{thebibliography}
\end{document}